\documentclass[pra,showpacs,floatfix,nofootinbib]{revtex4}
\usepackage{amsmath}
\usepackage{amssymb}
\usepackage{amsthm}
\usepackage{amscd}
\usepackage{amsfonts}
\usepackage[dvips]{graphicx}
\usepackage{latexsym}
\usepackage{mathrsfs}
\usepackage{bm}
\usepackage{subfigure}
\usepackage{color}

\usepackage{graphics}
\usepackage{graphicx}
\usepackage{pifont}
\usepackage{caption2}
\usepackage[a4paper,height=25cm,hmargin={2cm,2cm}]{geometry}
\makeatletter
\newcommand\figcaption{\def\@captype{figure}\caption}
\newcommand\tabcaption{\def\@captype{table}\caption}
\makeatother

\begin{document}
\title{A Mean-Field Analogue of the Hong-Ou-Mandel Experiment With Bright Solitons}

\author{Zhi-Yuan Sun}
\affiliation{Department of Physics, Technion-Israel Institute of 
Technology, Haifa, Israel}

\author{Panayotis\ G.\ Kevrekidis}
\affiliation{Department of Mathematics and Statistics, University of Massachusetts,
Amherst, MA, USA}

\author{Peter Kr\"uger}
\affiliation{Midlands Ultracold Atom Research Centre, School of Physics \& Astronomy, The University of Nottingham, Nottingham, UK}


\begin{abstract}
In the present work, we theoretically propose and numerically illustrate
a mean-field 
analogue of the Hong-Ou-Mandel experiment with bright solitons. More specifically,
we scatter two solitons off of each other (in our setup, the bright solitons play the role
 of a classical
analogue to the quantum photons of the original experiment), while the
role of the beam splitter is played by a repulsive Gaussian barrier.
In our classical scenario, distinguishability of the particles yields,
as expected, a $0.5$ split mass on either side. {\it Nevertheless}, for
very slight deviations from the completely symmetric scenario
a near-perfect transmission 
state can be constructed instead, very similarly to the
quantum mechanical output. We demonstrate this as a generic
feature under slight variations of the relative soliton speed, or of the
relative amplitude in a wide parametric regime. We also explore how
variations of the properties of the ``beam splitter'' (i.e., the
Gaussian barrier) affect this phenomenology.
\end{abstract}





\maketitle


\section{Introduction}

The Hong-Ou-Mandel (HOM) effect is, by now,  a well-established
experiment in quantum
mechanics that describes two-particle interference based on a pair of
indistinguishable photons \cite{Hong}. When those two identical
single-photon wave packets simultaneously enter a 50:50 beam splitter, one in each
input port, both always exit the splitter at the same output port,
although each photon has (on its own) 
a 50:50 possibility to exit either output
port. 
As a result of this effect, we
can test the degree of indistinguishability of two incoming photons
experimentally. While a {\em direct} measurement of both {\em quantum} particles (photons) exiting the beam splitter through the same port is normally not possible, the number of coincidence counts of the photons exiting the beam splitter through one exit port each dips to zero (the so-called HOM dip) in the case of perfect indistinguishability. Santori \emph{et al.} applied the HOM effect to
demonstrate the purity of a solid-state single-photon source~\cite{Santori}, 
while Beugnon \emph{et al.} experimentally considered
two atoms independently emitting a single photon to produce the
HOM effect~\cite{Beugnon}. From the point of view of applications, the HOM
effect has provided a mechanism for logic gates in linear optical
quantum computation \cite{Knill}. It is important here to stress the fact
that in the case where solitonic wave packets replace the photons, in contrast, a direct measurement does become possible and a {\em classical} mean-field description, as used in this work, can be employed.

As a generalization of the HOM effect, 
recent studies have been devoted to the interference of
massive particles. Lim and Beige have considered HOM experiments
with $N$ bosons or fermions passing simultaneously through a symmetric
Bell multiport beam splitter~\cite{Lim}. Longo \emph{et al.} have examined
the joint probability distribution (of finding both photons on the same
side) upon varying 
the properties of the beam-splitter~\cite{Longo}.
Lalo\"{e} and Mullin generalized the HOM effect for a large number of
particles by investigating quantum properties of a single beam splitter
\cite{Laloe}. In fact, Bose-Einstein condensates (BECs) at very low temperatures
provide a setup for studying an analogue to the HOM effect 
for massive particles.
Recently, Lewis-Swan and Kheruntsyan proposed the realization of the
HOM effect for massive particles
by using a collision of two BECs and a sequence of laser-induced Bragg pulses
as the splitter \cite{Lewis-Swan} (such an
experimental technique has been formerly
used to demonstrate the violation of the Cauchy-Schwarz inequality with matter
waves \cite{Kheruntsyan}). This has been further explored experimentally
very recently in a plasmonic setup using surface plasmon 
polaritons (SPPs)~\cite{spps}. Here the two photons were converted
into SPPs on a metal stripe waveguide and were subsequently made to
interact through a semi-transparent Bragg mirror. This resulted
in a coincidence count dip by $72 \%$ confirming the bosonic nature
of the SPPs and the HOM destructive quantum interference effect.

On the other hand, on the more ``classical'' side of matter waves,
bright solitary waves or solitons have been extensively studied in the
context of Bose-Einstein condensates~\cite{emergent}. Restricting
our consideration (for the present work) to attractive interactions
and bright solitary waves, we can note that both one~\cite{salomon}
and many~\cite{randy,randy1} (as a result of the 
modulational instability) such waves 
have been created in $^7$Li. More recently they have
also been produced in $^{85}$Rb~\cite{cornish}, but
furthermore the interactions between them and with barriers
have been explored~\cite{abdul,billamrev,Helm,Cuevas} both at the mean-field
and at the quantum mechanical~\cite{ruost,Gertjerenken}
level. Very recently 
experimental signatures  have also been reported
both for the case of interactions with barriers~\cite{Marchant}
and for those between bright solitons with different phases~\cite{hul_new}.
It should be noted here that while bright solitons~\cite{abdul,rev_ind}
and even their trains~\cite{usama} and collisions~\cite{parker_jpb}
constitute well-established themes in the BEC literature, interesting
variants thereof continue to emerge including bright solitons in
spin-orbit~\cite{djf,djf2} and exciton-polariton BECs~\cite{dimas,dimas2}.
Moreover, new experimental techniques for their production
(such as rf-evaporation for producing one or a pair of bright 
solitons)~\cite{kasevich} and their use in applications including
interferometry~\cite{robins} are being devised.

It is at the junction of these two exciting research themes that
the present study treads. The bright solitary waves
possess some quantum mechanical (or, more accurately, wave-like)
features, including a transmission and reflection from a 
repulsive Gaussian barrier (which we will consider
hereafter) as has been analyzed physically
in~\cite{Helm}, based to a large extent on the authoritative
earlier mathematical analysis of~\cite{Holmer}. 
However, they are not genuine quantum particles such as photons
superposed in Fock states. A manifestation of the latter 
feature is the fact that the HOM dip occurs when the photons
are perfectly indistinguishable. On the other hand, if 
two identical bright solitons enter the beam splitter perfectly 
symmetrically,  the result of their
collision will be a perfect splitting into an output state, with one soliton emerging from each port.

{\it Nevertheless}, what we argue here is that 
our mean-field treatment shows that even very slight
deviations of the bright solitons from perfect symmetry
(of the order of a few percent in the relative speed,
or in the relative amplitude) yield an output whereby
the bright solitons only emerge in one of the two (controllably
so, depending on the sign of the asymmetry) ports i.e.,
as an analog of the $|2,0 \rangle$ or the $|0,2 \rangle$
state. Indeed, it is this analogy with the HOM feature
of revealing both ``particles'' (in our case, the bright solitons)
 on one side which constitutes
a remarkable feature that arises over a
wide range of values of both the soliton and the barrier
parameters and as such can be considered as {\it generic}.
We argue that this phenomenon cannot purely emerge from the interaction
of the bright solitons with the barrier but must stem critically from their
pairwise interaction during their ``coincidence'' at the barrier. 
These deviations in soliton parameters (that give rise to these
asymmetries) are so weak that they can very straightforwardly
arise due to the imperfect preparation of the colliding bright solitons.
Thus, the constructive/destructive
interference of our (very) weakly asymmetric bright solitons 
at the barrier is responsible
for their emergence on one or the other port in this setting.

Our presentation is structured as follows. In section II, we briefly
discuss the theoretical setup. Section III contains our numerical
results and a discussion of the variation of our phenomenology
over regimes of speed and amplitude variations of the bright solitons
and amplitude and width variations of the barrier.
Finally, section IV contains a summary of our findings and
a number of suggestions for potential future studies.

\section{Theoretical Setup}

We set two bright
solitons of BECs to collide at a narrow barrier of Gaussian form,
which is viewed as the analog of the beam splitter. 
We should note here that this kind of setup has been
extensively studied recently. The thorough exposition and analysis
of~\cite{Helm} explored the outcome of the collision of two 
identical bright solitons, examining chiefly --and even semi-analytically,
following the work of~\cite{Holmer}-- the role of asymmetries
in the phase between the bright solitons (section IV.A therein). The
issue of asymmetry in bright soliton amplitudes was briefly discussed
as well (in section IV.B therein) without a special focus, to the
best of our understanding, on the phenomenology reported here.
In particular, the central question (here) 
of the outcome of uneven bright soliton velocities was not considered
in~\cite{Helm}. On the other hand, the work of~\cite{Cuevas}
reports observations very similar to the focus of the
present work (see e.g. Fig.~8 therein and in particular the
evolution of the two bright solitons colliding after $t=0.1$s in the top panel).
Nevertheless, as this
work concerned the collision of a single bright soliton with a barrier
(where the interaction of two bright solitons with the barrier was only
a secondary effect, due to the ``return interaction'' of the
two splinters), this feature was not examined systematically,
although glimpses of it can be inferred by the second reflection
coefficient plots of Fig. 5 in~\cite{Cuevas}.

To simplify the relevant context, we examine 
the collision phenomenology
in the setting of
the normalized quasi-1D Gross-Pitaevskii (GP) equation with 
attractive interactions:
\begin{equation}
i \frac{\partial \psi(x,t)}{\partial t} = \left[ - \frac{1}{2}
\frac{\partial^2}{\partial x^2} + \frac{q}{\sigma \sqrt{2\pi}}
e^{-x^2/(2 \sigma^2)} - |\psi(x,t)|^2 \right] \psi(x,t)~,\label{1}
\end{equation}
where $\psi(x,t)$ is the dimensionless wave function with normalized
temporal and spatial coordinates $t$ and $x$;
see e.g.~\cite{Helm} for a discussion of the relevant
units. The dimensionless form of the equation has been systematically
derived e.g. in~\cite{emergent} (see Ch. 1 therein, while Ch. 2
is specifically dedicated to bright solitons).
While, as discussed
in~\cite{Cuevas}, quantitative features may be expected to differ
in the 3D case (and the latter may also differ even qualitatively
for near-critical atom numbers in the vicinity of the collapse
threshold), qualitative features of the 1D GPE can be fairly accurate
for a wide range of atom numbers (up to $\approx 0.7 N_c$, where
$N_c$ is the critical atom number); 
see e.g. the comparison in Fig.~1 therein.

The Gaussian barrier
has a normalized width $\sigma$ and strength $q$. 
Our initial condition in the present work involves two
oppositely moving bright solitons of the form:
\begin{equation}
\psi(x,t=0) = k_1 \textrm{sech} [k_1 (x+x_1)] e^{i v_1 x} +  k_2
\textrm{sech} [k_2 (x-x_2)] e^{-i (v_2 x + \Delta)}~.\label{2}
\end{equation}
For  sufficiently large values of $x_1$ and $x_2$ ($x_{1,2}>0$), 
Eq.~(\ref{2}) 
approximately represents a pair 
of two bright solitons located at  $-x_1$ and $x_2$,
with amplitudes $k_1$ and $k_2$, oppositely moving velocities $v_1$
and $-v_2$ ($v_{1,2}>0$), and with a relative phase $\Delta$. 
Although this ansatz is reminiscent of the one used
in~\cite{Helm}, contrary to that study here we will generally
not utilize the phase difference, setting it to $\Delta=0$, unless
indicated otherwise; a relevant brief comment on its role
is included in the next section.
Instead, a critical distinguishing feature of the present
work will be that we will be relying on slight asymmetries
of the propagation characteristics of the solitons such
as their speeds or their amplitudes/inverse widths in order
to achieve our mean-field analogue of the HOM effect.
Note that the limiting case of a $\delta$-shaped barrier 
($\sigma \rightarrow 0$ for the Gaussian barrier) has been 
treated analytically 
by 
Holmer \emph{et al.}~\cite{Holmer} for a single bright soliton.

As our mean-field experiment with the two bright solitons playing the role
of (partially classical) analogs of the photons in the original
HOM setting, 
and the barrier acting as the beam splitter,
we arrange for the two matter-wave
bright solitons located at $-x_1$ and $x_2$ to collide exactly
at the center of the Gaussian barrier, which requires $x_1/v_1 = x_2
/v_2$. The essential point is to control a slight difference between
the velocities $v_{1,2}$. In our setup, we ensure that 
$\left|\frac{v_2-v_1}{v_1}\right|
\leq0.1$.  As numerical diagnostics of the ``mass'' (i.e., atom
number fraction) that emerges on each side as a result of this
experiment, we compute two
normalized integral quantities:
\begin{equation}
E_+ = \frac{\int_0^{+\infty}|\psi|^2 dx}{\int_{-\infty}^{+\infty}|\psi|^2 dx}~,~~~~~~
E_- = \frac{\int_{-\infty}^0|\psi|^2 dx}{\int_{-\infty}^{+\infty}|\psi|^2 dx}~.\label{3}
\end{equation}
It is then well known (and easily understood by symmetry) 
that for $v_1=v_2$ and
other parameters chosen the same for both incoming bright solitons, it will be
true by construction
that $E_+=E_-=0.5$. We now turn to the case of unequal velocities
and amplitudes in our computations presented below.

\section{Numerical Results and Discussion}

In the case of unequal
velocities, we illustrate typical realizations in Figs.~\ref{A} (the numerical simulation is performed
using a 4th-order Runge-Kutta algorithm
in time,  2nd-order centered difference in space 
scheme). As seen (and in line with observations such as the secondary
collision of Fig.~7 in~\cite{Cuevas}), the small difference ($\approx 2\%$)
in the bright soliton velocities
induces a {\it dramatic deviation} from the above mentioned 
equal velocity result. In other words, 
envisioning two detectors located on
the bright soliton moving paths after collision, we observe that
essentially a sole, double in mass solitary wave packet will
be found at one detector, with nearly no mass collected at the other one
(the partition is nearly $99\%$ and $1\%$ in the example shown in 
Fig.~\ref{A}). 
I.e., in the HOM Fock space language 
a $|2,0\rangle$ or $|0,2\rangle$ state (cf. Fig.~\ref{A})
is recovered rather than a coincidence
count of $|1,1 \rangle$, in some sense similarly to the genuine quantum
particle result.
 Unlike the quantum case, the {\em entanlged} state $[|2,0\rangle + 
|0,2\rangle]/\sqrt{2}$ is not what is found here. Instead, the slight asymmetry 
determines uniquely on which output port 
of the beam splitter the double mass bright soliton is found, thus 
replacing the decoherence process of the quantum case.

Fig.~\ref{B} examines the role of the difference between $v_1$ and $v_2$
by fixing $v_1=1$, and varying $v_2$ in a small range $0.9$ to 
$1.1$. The simulation
results as illustrated by the quantities 
$E_{+,-}$ are shown in the figure. We
see that a peak value occurs when $v_2/v_1$ approaches 1 from lower
values,
which means most of the bright soliton energy is found in one detector. 
The situation is 
symmetric as $v_2/v_1$ approaches  1 from higher values, and the energy 
(peaking again around $99\%$) is
detected in the other output port this time. The remarkable feature
is that for such a dramatic effect, miniscule deviations of the order of
only $1$-$2\%$ are sufficient. As observed in Fig.~\ref{B},
the asymmetry is maintained above $70 \% - 30 \%$ nearly throughout
our parametric interval of observation. 
The tiny region near $v_1=v_2$  
 is an exception that is explainable by the setup of our model (see above). Experiments are eventually expected to show whether this result prevails empirically. If it does, the narrowness of the feature could open the path to using bright soliton collisions for sensitive  measurements of minute external forces modifying the effective travel path length. 
Aside from this narrow region, it can be observed that 
a situation most closely reminiscent of the original HOM output
is best replicated when the bright solitons have a roughly $50 \%$ transmission
and reflection probabilty from the barrier; this will be further
illustrated in what follows.

We should note here that for a single bright soliton, the numerical results of Helm 
\emph{et al.}~\cite{Helm} suggest
that, in the approximate range $0.5 \leq v \leq 2$ and $\sigma \leq
0.28$, the bright soliton should tunnel through the Gaussian barrier
instead of classically passing through. As a result, the authors
conclude that the following
relation is satisfied within this tunneling regime,
\begin{equation}
\frac{1}{2} v^2 \ll \frac{q}{\sigma \sqrt{2 \pi}}~.\label{4}
\end{equation}
This translates into $v \ll 2.8$ for $q=1$ and $\sigma=0.1$ to achieve the
tunneling regime. 

In Figs.~\ref{C} we illustrate how the effect of slight
velocity difference explored above 
(in analogy with HOM) is modified for faster, as well as more
slowly-moving bright solitons within this tunneling regime. 
Several features can be distinguished here. Firstly, there is
an ``optimum'' as regards the mass asymmetry involved in the
process. This appears to arise when $v_1 \approx q$. 
This, in turn, suggests a nearly 50:50 beam splitter, given
the expressions for transmission and reflection 
from a $\delta$ barrier [cf. e.g. Eqs. (7)-(10) and associated
discussion in~\cite{Helm}].
Secondly,
the location of the maximal asymmetry is monotonically approaching
$v_2/v_1 =1$, as $v_1$ is decreasing. Nevertheless, and while for
large $v_1$, the phenomenon is more pronounced, with a rapid
decrease of asymmetry as $v_2/v_1$ deviates more significantly
from unity, the opposite is true in the slow case. For small
$v_1$ (slow bright solitons), although the peak approaches $v_2/v_1=1$,
the curve also flattens and becomes nearly insensitive to the exact
value of $v_2/v_1$ and the corresponding asymmetry is far less
pronounced. From the above, we infer that this asymmetry is most
evident when $v_1$ (and $v_2$) are near $q$ and the relevant
mass peak in that case is very proximal to unity, while it occurs
only within a few percent of the $v_1=v_2$ limit.
It should be added here that in line with the discussion of~\cite{Helm},
such an asymmetry cannot be justified by a brief (rapid) interaction of the
solitary waves with the barrier. Exploring the formulation
of~\cite{Holmer} in the same way as is done in Sec. IV.A of~\cite{Helm},
it can be shown that the relative mass (the factors of $|P_-|^2$ and
$|P_+|^2$) smoothly deviates from $1/2$ by only a few percent and hence
cannot justify the asymmetry we observed. It must thus be that
this outcome is due to the interaction of the solitary waves
with each other {\em and} with the barrier.
For vanishing inter-soliton interaction, each bright soliton would individually 
interact with the barrier, so that slight asymmetries would only cause 
slight deviations from a 50:50 splitting, always resulting in close to equal 
population at the two outputs of the beam splitter in the presented setup. 
Conversely, for a collision event of two bright solitons in the absence 
(or far away from) the barrier, the interactions would be far too weak to 
cause a significant deviation from a 50:50 split (for identical
or near-identical bright solitons), i.e. the bright solitons 
would essentially pass through each other, as they would
constitute exact bright solitonic solutions of the integrable
1D homogeneous GP equation.

Although in all other results reported in the manuscript, the
relative phase of the bright solitary waves is initially chosen
to be $\Delta=0$, given the difficulties in experimentally controling
such a phase it is, arguably, of relevance to explore the role of
$\Delta$ in potentially affecting the above results. This is 
examined in Fig.~\ref{C_extra}. In the top panel of the figure,
 it can be seen that indeed
variations in the original phase difference $\Delta$ will shift
the speed ratio $v_2/v_1$ of optimal induced asymmetry (i.e., 
of maximal $E_+$). Nevertheless, as illustrated in the bottom panel,
the variation is only linear and its small slope ensures that the 
phenomenology presented above (and below) will be relevant even
in the presence of nontrivial phase differences $\Delta$.

We now consider different variants of the relevant phenomenology.
If, for instance, the two bright solitons collide a bit further
away from the center of the Gaussian barrier, the effect observed above
still persists as in Fig.~\ref{C}. We consider two bright solitons
starting at $\pm x_0$, and they collide at the approximate position
$x = x_0 \left( \frac{1-v_2/v_1}{1+v_2/v_1} \right)$. This situation is
similar to the case when $v_2/v_1\rightarrow1$, although
the maximal asymmetry may occur, e.g., 
as $v_2/v_1\rightarrow0.9$. With our parameters
($x_0 = 20$), the collision position is $x\approx1.05$ when $v_2/v_1=0.9$,
and parts of the colliding bright solitons are still within the scale of the
Gaussian barrier (for $k_1=k_2=1$). Our simulation shows that the difference
on $E_{+}$ is small as $v_2/v_1\rightarrow0.9$, and smaller especially for
the slow bright solitons.

A particularly interesting variation of the theme is that asymmetries in
bright soliton amplitudes/width may also be used to produce a complete asymmetry
(in either direction) of the collisional outcome.
This is illustrated by varying $k_1$ while keeping $k_2$ fixed and 
$\left|\frac{k_2-k_1}{k_1}\right|
\leq0.1$ in our simulations shown in Fig.~\ref{D}. In panel (a), 
we give a group of results for
the bright soliton collision with $v_1=v_2=1$. The variation of amplitude
(or inverse width)
distribution with $k_2/k_1$ is similar to that with $v_2/v_1$ after collision.
However, for bright solitons with smaller amplitude and larger width (our
simulations are based on bright solitons with amplitude comparable to the barrier
height, and width much larger than the barrier width), the curve $E_-$ is
more flat and the effect is considerably less pronounced,
especially so for slower bright solitons; cf. Fig.~\ref{D}(b).

Another key element in the analysis of this HOM-type phenomenology
is the role of the ``beam splitter'' i.e., how the observations are
modified by varying the parameters of the barrier.
To analyze that, we fix $v_1=1$, and
give two simulation results in Fig.~\ref{E} with variation of the
barrier strength and width. Eq.~(\ref{4}) suggests that when
$\sigma\ll0.8$ (for $q=1$), the bright soliton steps into the  tunneling
regime. Fig.~\ref{E}(a) shows that the curve $E_+$ is gradually
approaching a constant value of 0.5 as $\sigma\geq1$, which can be seen as 
being outside the tunneling regime. On the other hand, as 
$\sigma$ approaches 0.1, the 
mass asymmetry is maximized, reaching  $\approx 98\%$ in one detector.
However, if we further decrease the width, the maximal asymmetry decreases
indicating a non-monotonic dependence. 
The situation is similar (and again non-monotonic) with variation of $q$. When
$q\gg0.13$ ($\sigma=0.1$), the bright soliton is considered to be in the
tunneling regime. Fig.~\ref{E}(b) shows that, when $q<0.1$, the curve
is gradually approaching again to the constant value of 0.5. 
The maximum asymmetry of our HOM-like detection can be
observed as $q$ approaches to 1 (i.e., to $v_1$). 
Finally, further increase of $q$ beyond the above maximum yield value
of $v_1$, leads anew to a less pronounced phenomenology and to a flattening
of the relevant curve.

To examine the role of potential fluctuations (and asymmetries)
of the beam splitter itself (rather than of the bright solitons), we have also
examined the possibility of adding
random noise on top of the Gaussian barrier in
Eq.~(\ref{1}). This was implemented as
$V(x) = \frac{q}{\sigma \sqrt{2\pi}}[1+ \eta \varepsilon(x)]
e^{-\frac{x^2}{2 \sigma^2}}~,$
where $\eta$ is the noise strength, and $\varepsilon(x)$ is a random
function with uniformly distributed random values in
$[-1,1]$. For the weak noise ($\eta=0.1$), we have performed
simulations in the case of faster ($v_1=1$, 12 realizations) and slower
($v_1=0.2$, 6 realizations) solitary waves, respectively. 
From our observations, we conclude that our findings are
only weakly affected by this type of 
slight variations/asymmetries of the barrier.
It is instead chiefly the weak asymmetry 
of the solitary waves, in the appropriate parametric
regime as per the discussion above, that is responsible
for the observed phenomenology.

Lastly, in order to offer a glimpse of theoretical 
insight towards the numerical observations presented herein,
we propose the following heuristic argument, which we have
tested to be valid for large speeds of the incoming
solitary waves. In this case,
our observations here appear to capture the slight asymmetry
between the incoming solitary waves building a relative
phase difference, $\Delta$, between them. The latter, in turn, and
in agreement with the arguments of~\cite{Helm} would
provide a maximal asymmetry in the collisional output
if it assumes the value of $\Delta=\pi/2$. However, we  assume
that our solitary waves are  
fast enough, then their accumulated
phase difference (over the time $t=x_1/v_1$ needed to reach
the barrier collision point) is 
$\Delta=(v_1^2-v_2^2) t/2$, as stems from the expression of
integrable bright solitons of the GP equation.
Setting these two expressions for $\Delta$ equal, we retrieve
an estimation of the optimal asymmetry in the form of:
$(v_2/v_1)^2=1 - \pi/ (x_1 v_1)$. While this is a relatively
simplistic calculation, it qualitatively agrees with 
our numerical computations. In particular, it has prompted us
  to examine the dependence of the 
point of optimally asymmetric output 
on the starting bright soliton location(s), i.e., on $x_1$.
An example of this is shown in Fig.~\ref{fignew}. Indeed, the latter clearly
displays the existence of a monotonically decreasing asymmetry trend 
in the location of the optimum as $x_1$ is increased. Moreover,
for the large speeds used in this example, the
second panel of the figure illustrates that the agreement
for the prediction of the relevant optimal asymmetry is not merely
qualitative but also quantitative.

\section{Conclusions and Future Challenges}

In the present work, we have proposed a mean-field experiment
with bright solitons interacting with each other, at a Gaussian
barrier, in a way
analogous to the Hong-Ou-Mandel experiment with photons (or
more generally bosons). In this analogy, 
the role of the photons is played by
the bright solitons,  and that of the beam splitter by a Gaussian barrier.
Our findings are rather unexpected in many ways. In the limit of 
perfect symmetry, the output result is, as expected classically
(and based on symmetry), an even mass split between the two output
ports. On the other hand, in a way somewhat reminiscent
of the quantum mechanical analogue, for weak deviations from
``indistinguishability'', the mean-field treatment of the bright solitons leads to a strongly
asymmetric result, one of a nearly perfect $|2,0\rangle$ 
or $|0,2\rangle$ state. We have quantified this effect
and have illustrated its occurrence both for (weak) asymmetries
of incoming bright soliton velocities or even for ones of bright soliton
amplitudes (or inverse widths). This phenomenology has been
quantified over variations of parameters of the barrier (such
as its strength and inverse width) and relevant optima
have been revealed (e.g. when the strength of barrier
is nearly comparable to the incoming bright soliton velocities, i.e.,
in the nearly 50:50 beam-splitter regime).

These results pave the way for a considerable number 
of additional investigations in this field. On the one hand,
from a more mathematical perspective, it becomes especially relevant
to consider the appropriate extension of the work of~\cite{Holmer}
and a potentially deeper/more quantitative 
understanding of the role of asymmetries
in multi-soliton collisions (with a barrier). On the other hand,
from a physical perspective, it would be especially interesting
to explore whether different wave entities feature similar
behavior upon their interactions with barriers.
For instance, in the context of repulsive condensates,
it would be interesting to examine whether weakly asymmetric
dark solitons~\cite{djf3}  or perhaps even asymmetric vortices~\cite{fetter1}
may yield similar features in their pairwise interactions with barriers.
The fact that such interactions in both one-
and even in multi-component
settings have recently started to be  considered~\cite{Alvarez} suggests
the relevance of such studies. On the other hand, multi-component variants
of the problem 
would be worthwhile to explore even in the self-attractive
case, 
where in addition to the potential asymmetry in output ports,
further asymmetries between components can be envisioned.

Another potentially rather challenging direction may
involve the recently analyzed analogy between the Lieb-Liniger 
exactly solvable model 1D solutions and their mean-field bright solitonic
counterparts in the larger atom number limit~\cite{bettina} to explore the
question discussed herein for structures involving different
atom numbers. As this control parameter decreases, we can gradually
progress from the mean-field limit of the present work to the
quantum mechanical realm of the Lieb-Liniger model and examine
how the latter may modify the presently reported phenomenology.
These topics are currently under consideration and will be presented
in future work.

\vspace{0.5cm}

\textbf{Acknowledgments.}
Z.-Y.S. acknowledges support in part at the Technion by a fellowship
of the Israel Council for Higher Education. Z.-Y.S. also thanks 
the Department of Mathematics and Statistics at UMass Amherst
for the hospitality during his visit there.
P.G.K. acknowledges support from the National Science Foundation
under grants CMMI-1000337, DMS-1312856, from the ERC and
FP7-People under grant
IRSES-605096 from
the Binational
(US-Israel) Science Foundation through grant 2010239,
and from the US-AFOSR under grant FA9550-12-10332. 
P.G.K. is also grateful to J. Stockhofe for numerous
informative discussions on the HOM experiment
and related themes, as well as to J. Cuevas for 
discussions at the early stages of this project. 
P.K. acknowledges support by EPSRC 
(grant EP/I017828/1) and 
the European Commission (Grant No. FP7-ICT-601180). 
P.G.K. and P.K. also acknowledge
the hospitality of the Synthetic Quantum Systems group 
and of Markus Oberthaler
at the Kirchhoff Institute for Physics (KIP) at the University
of Heidelberg, as well as that of the Center for Optical 
Quantum Technologies (ZOQ) and of Peter Schmelcher at
the University of Hamburg.



\newpage
\noindent\textbf{Figures and Captions}
\vspace{1cm}
\\[\intextsep]
\begin{minipage}{\textwidth}
\renewcommand{\captionlabeldelim}{.\,}
\renewcommand{\figurename}{Figs.\,}
\renewcommand{\captionfont}{}
\renewcommand{\captionlabelfont}{}
\vspace{-0.6cm}\centering
\includegraphics[scale=0.6]{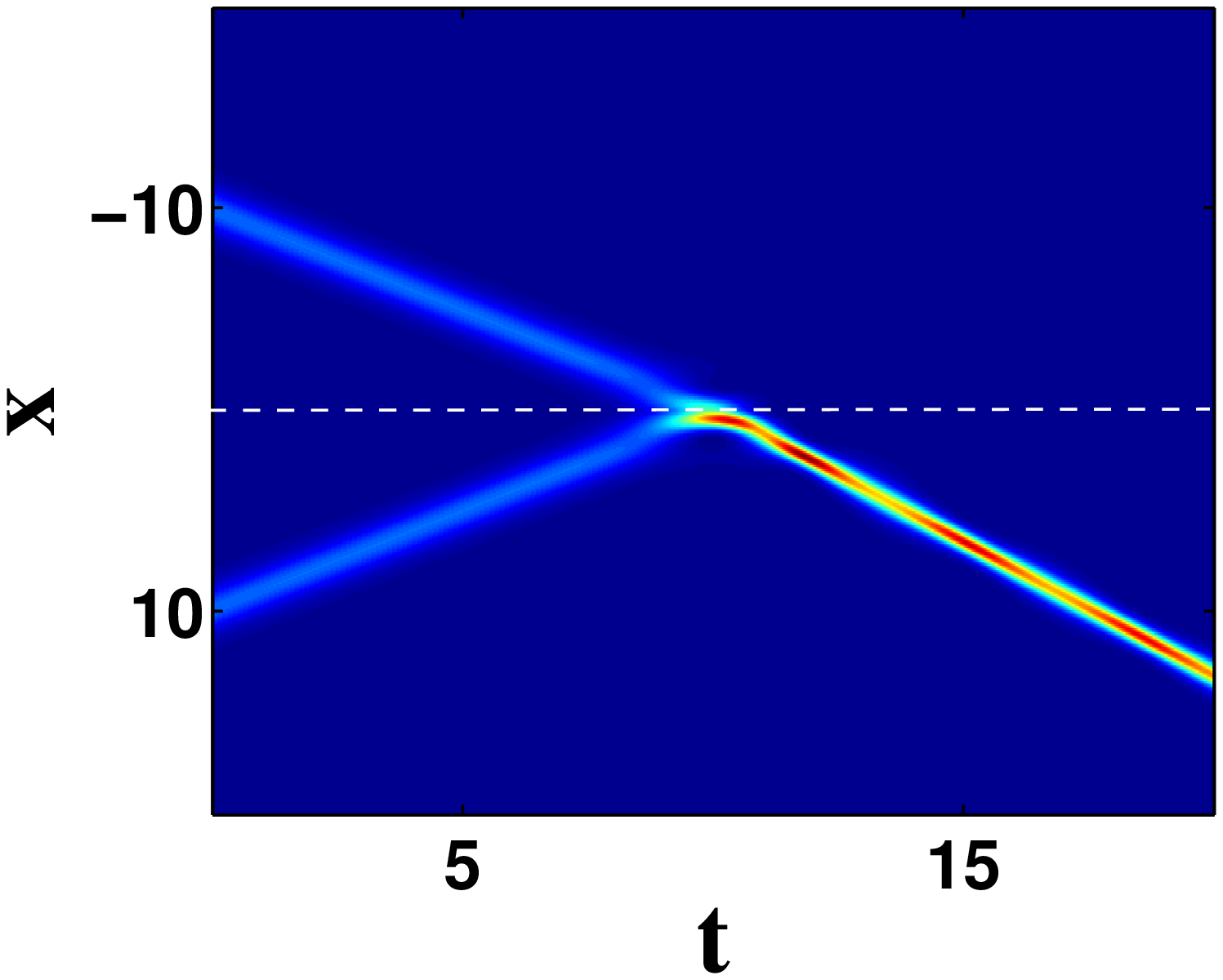}
\vspace{0cm} {\center\footnotesize\hspace{0cm}\textbf{(a)}}\\
\includegraphics[scale=0.6]{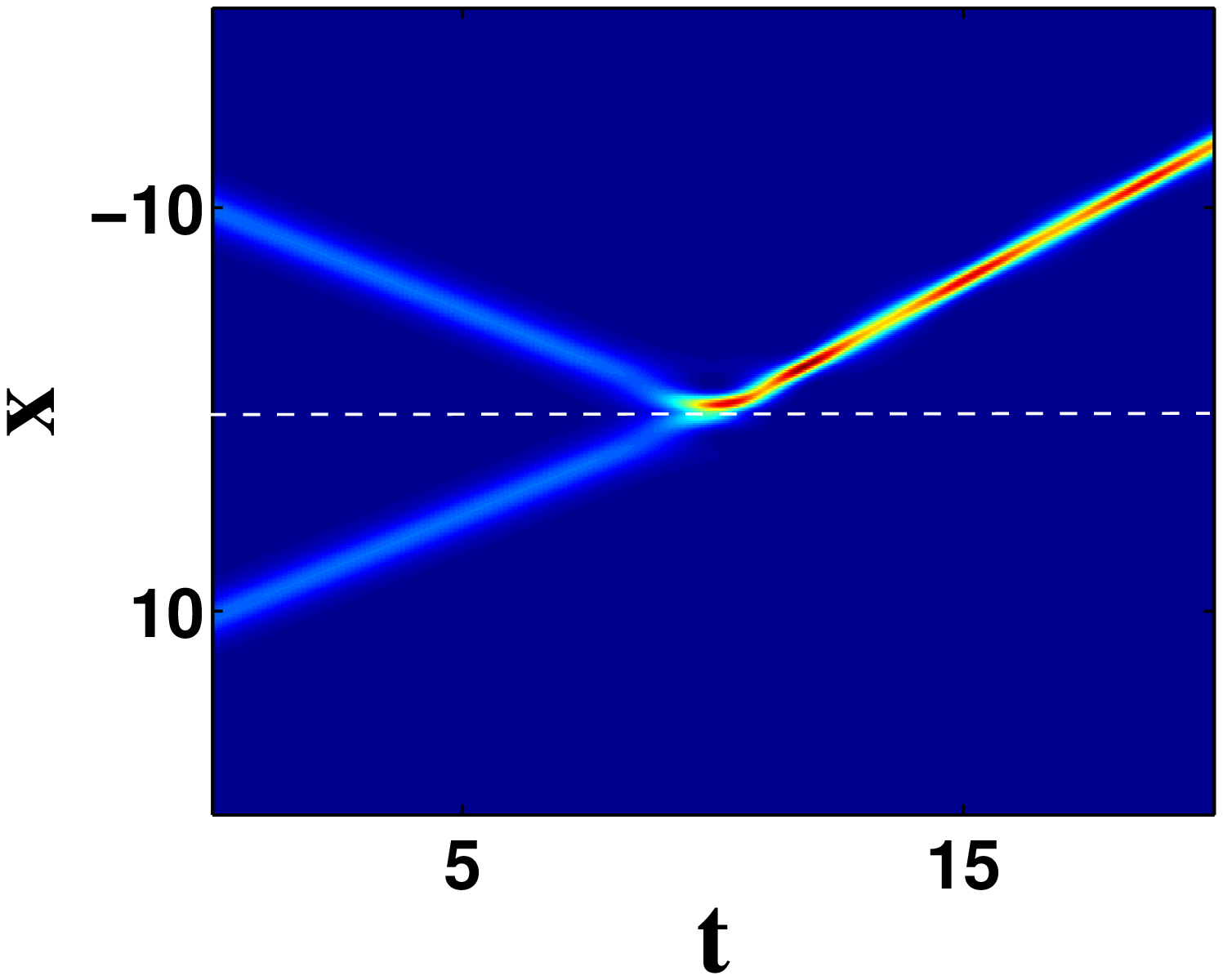}
\vspace{0cm} {\center\footnotesize\hspace{0cm}\textbf{(b)}}\\
\figcaption{(Color Online) Numerical simulation of the HOM analogous effect for
a two (slightly asymmetric) 
matter-wave bright soliton collision at the Gaussian potential
barrier. The parameters are chosen as $q=1$, $\sigma=0.1$,
$k_1=k_2=1$, and $\Delta=0$. (a) $v_1=1.00$, $v_2=0.98$, and
$x_1/v_1=10$; (b) $v_1=1.00$, $v_2=1.02$, and
$x_1/v_1=10$.  }\label{A}
\end{minipage}
\\[\intextsep]
\\[\intextsep]
\begin{minipage}{\textwidth}
\renewcommand{\captionlabeldelim}{.\,}
\renewcommand{\figurename}{Figs.\,}
\renewcommand{\captionfont}{}
\renewcommand{\captionlabelfont}{}
\vspace{2cm}\centering
\includegraphics[scale=0.55]{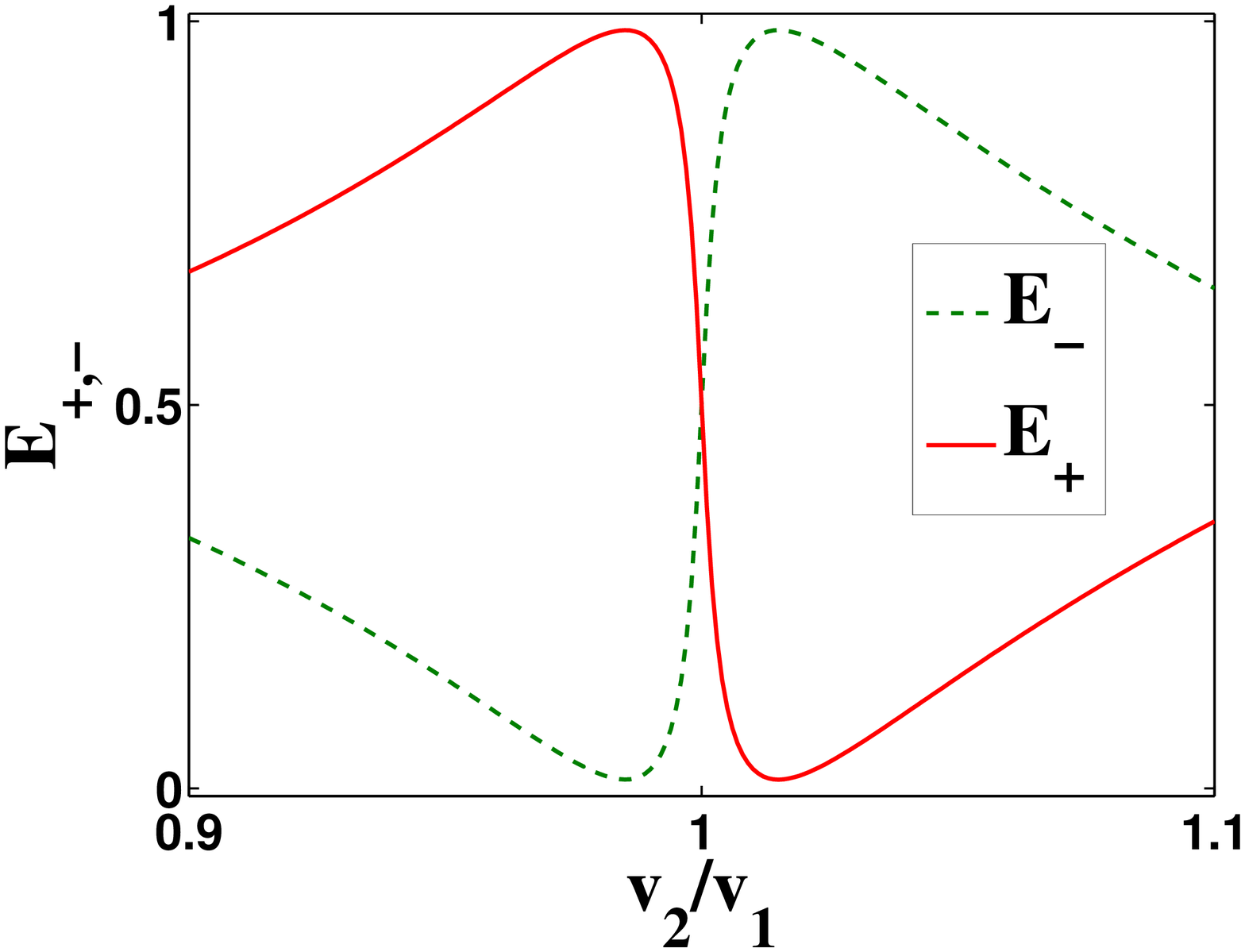}
\figcaption{(Color Online) Plots of $E_{+,-}$ (see Eq.~(\ref{3})) as 
a function of $v_2/v_1$ varying from 0.9 to
1.1 ($v_1=1$). The parameters are chosen as $q=1$, $\sigma=0.1$,
$k_1=k_2=1$, $\Delta=0$, and $x_1/v_1 = x_2/v_2 = 20$. }\label{B}
\end{minipage}
\\[\intextsep]
\\[\intextsep]
\begin{minipage}{\textwidth}
\renewcommand{\captionlabeldelim}{.\,}
\renewcommand{\figurename}{Figs.\,}
\renewcommand{\captionfont}{}
\renewcommand{\captionlabelfont}{}
\vspace{-1cm}\centering
\includegraphics[scale=0.55]{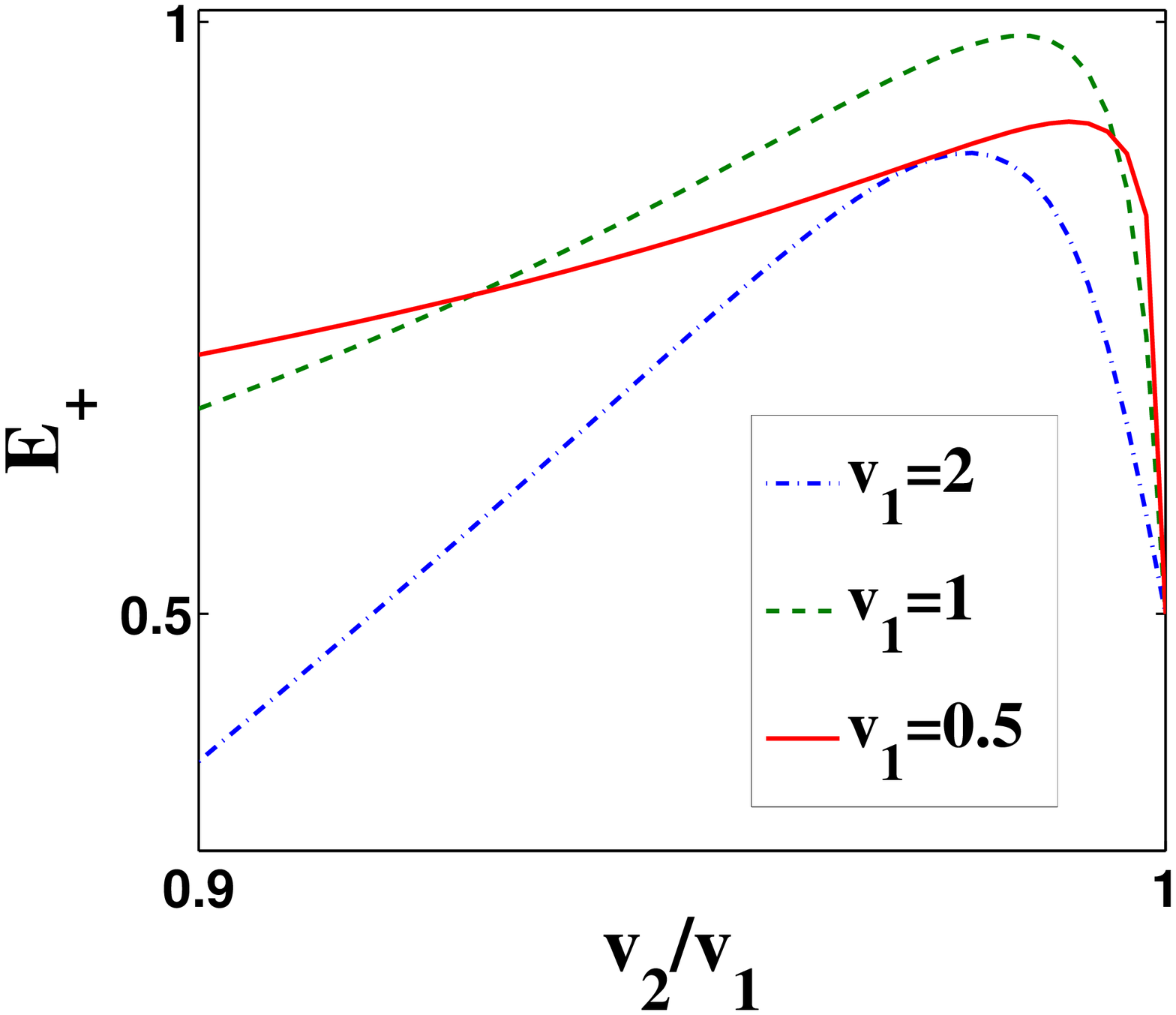}
\vspace{0cm} {\center\footnotesize\hspace{0cm}\textbf{(a)}}\\
\includegraphics[scale=0.55]{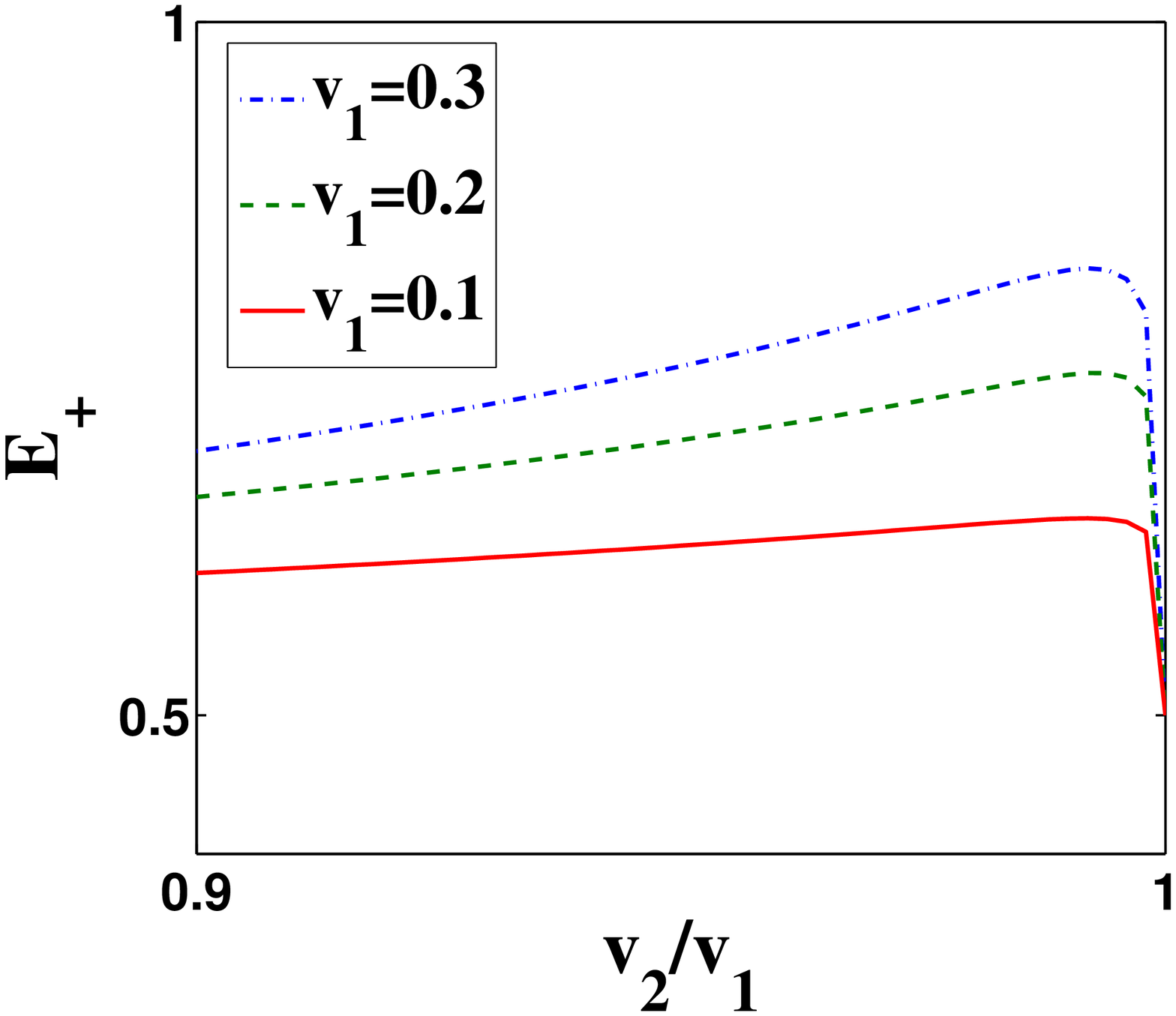}
\vspace{0cm} {\center\footnotesize\hspace{0cm}\textbf{(b)}}\\
\figcaption{(Color Online) Plots of $E_+$ (see Eq.~(\ref{3})) 
with $v_2/v_1$ varying from 0.9 to 1.0
for two groups of three values of 
$v_1$. (a) $v_1= 0.5$, $1.0$, and $2.0$; (b)
$v_1=0.1$, $0.2$, and $0.3$. The parameters are chosen as $q=1$,
$\sigma=0.1$, $k_1=k_2=1$, $\Delta=0$, and $x_1/v_1 = x_2/v_2 = 20$.
}\label{C}
\end{minipage}
\\[\intextsep]
\\[\intextsep]

\begin{minipage}{\textwidth}
\renewcommand{\captionlabeldelim}{.\,}
\renewcommand{\figurename}{Figs.\,}
\renewcommand{\captionfont}{}
\renewcommand{\captionlabelfont}{}
\vspace{-1cm}\centering
\includegraphics[scale=0.6]{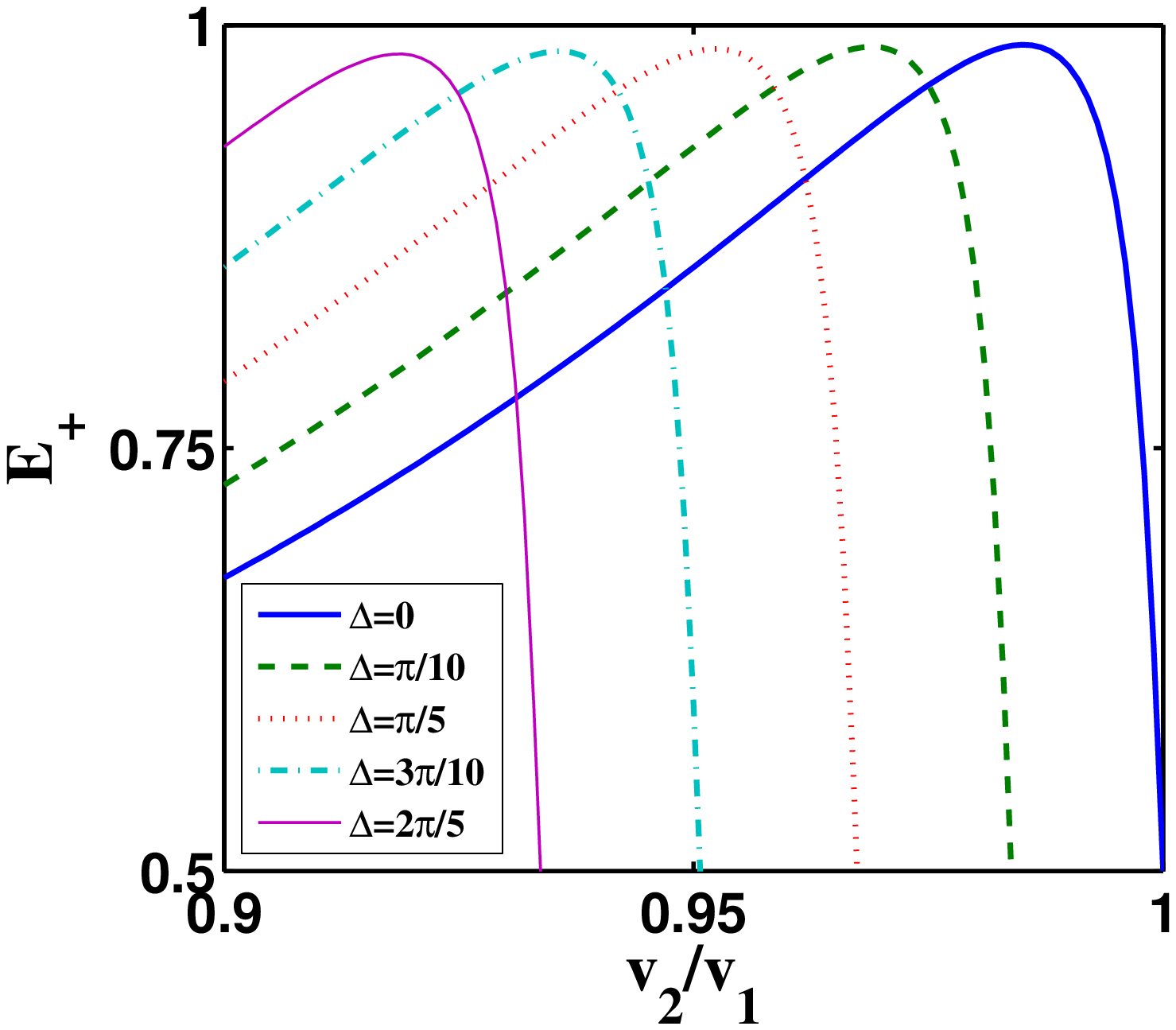}
\vspace{0cm} {\center\footnotesize\hspace{0cm}\textbf{(a)}}\\
\includegraphics[scale=0.6]{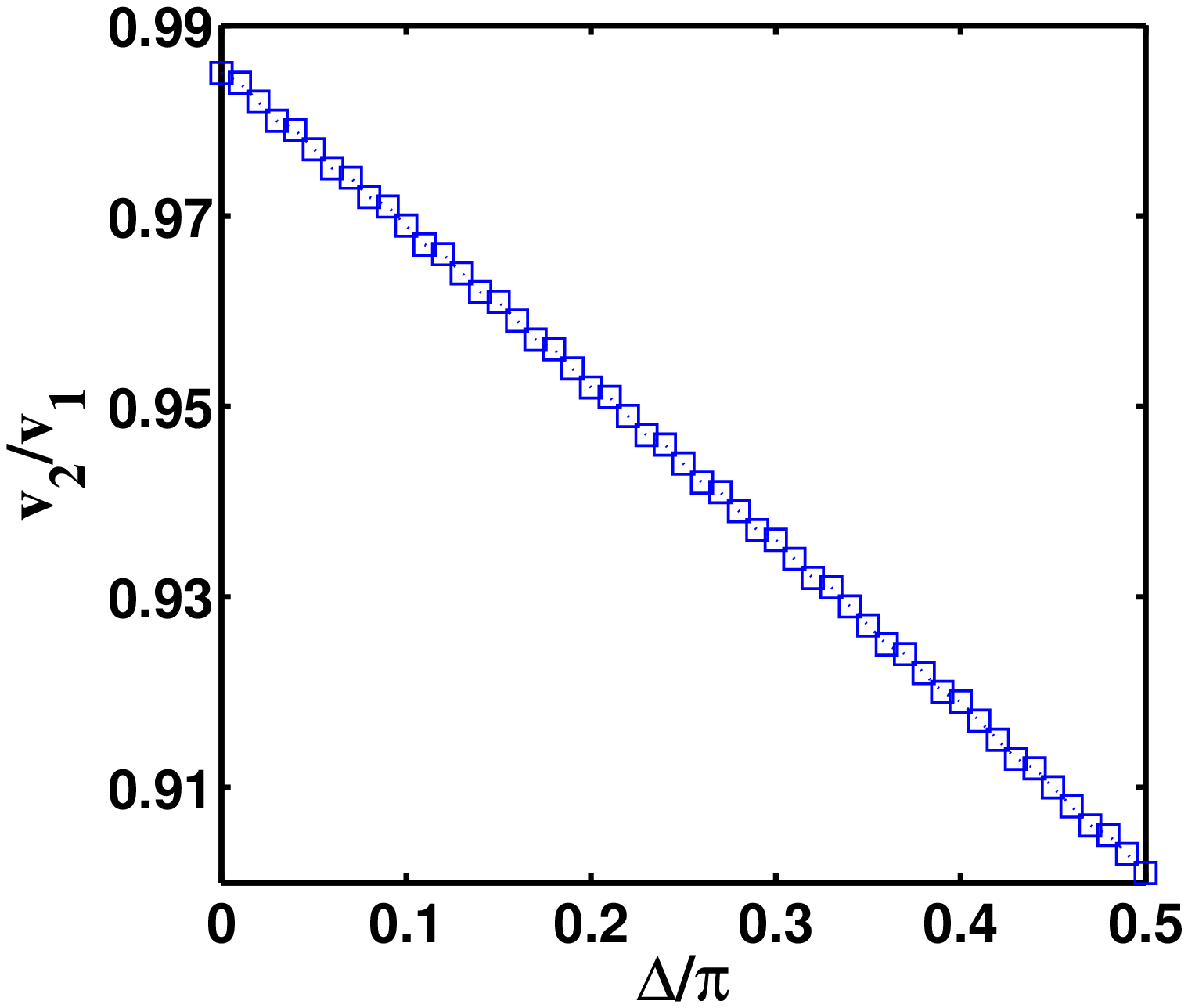}
\vspace{0cm} {\center\footnotesize\hspace{0cm}\textbf{(b)}}\\
\figcaption{(Color Online) The top panel shows the variation of 
the fraction $E_+$ as a function of the velocity for
different relative phases between the bright solitons. The
bottom panel shows the shift of the optimal point (of maximal asymmetry)
as a function of the relative phase $\Delta$. 
The other parameters are chosen as $q=1$, $\sigma=0.1$, 
$k_1=k_2=1$, $v_1=1$, and 
$x_1/v_1=x_2/v_2=20$.}
\label{C_extra}
\end{minipage}
\\[\intextsep]
\\[\intextsep]

\begin{minipage}{\textwidth}
\renewcommand{\captionlabeldelim}{.\,}
\renewcommand{\figurename}{Figs.\,}
\renewcommand{\captionfont}{}
\renewcommand{\captionlabelfont}{}
\vspace{-1cm}\centering
\includegraphics[scale=0.6]{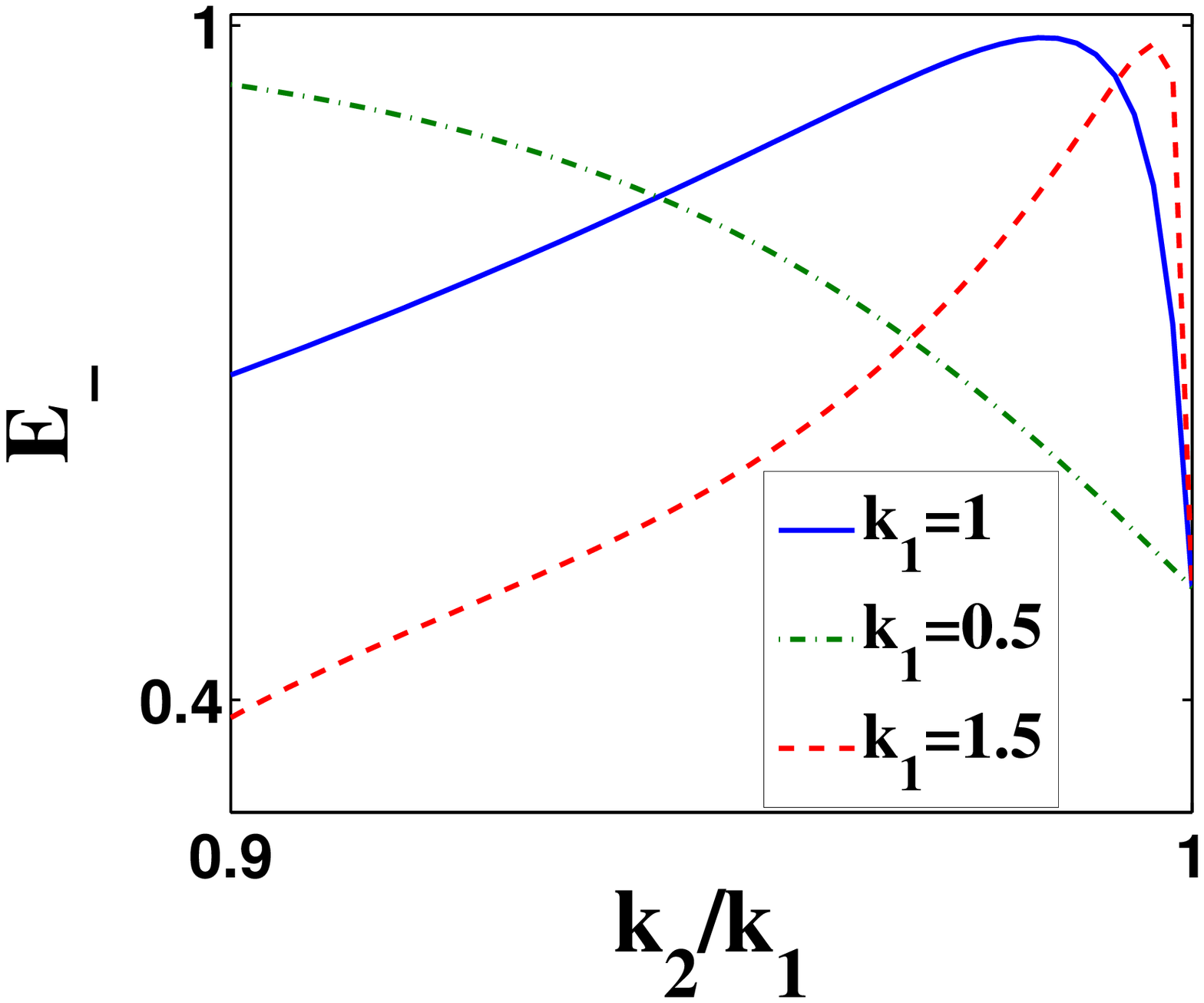}
\vspace{0cm} {\center\footnotesize\hspace{0cm}\textbf{(a)}}\\
\includegraphics[scale=0.6]{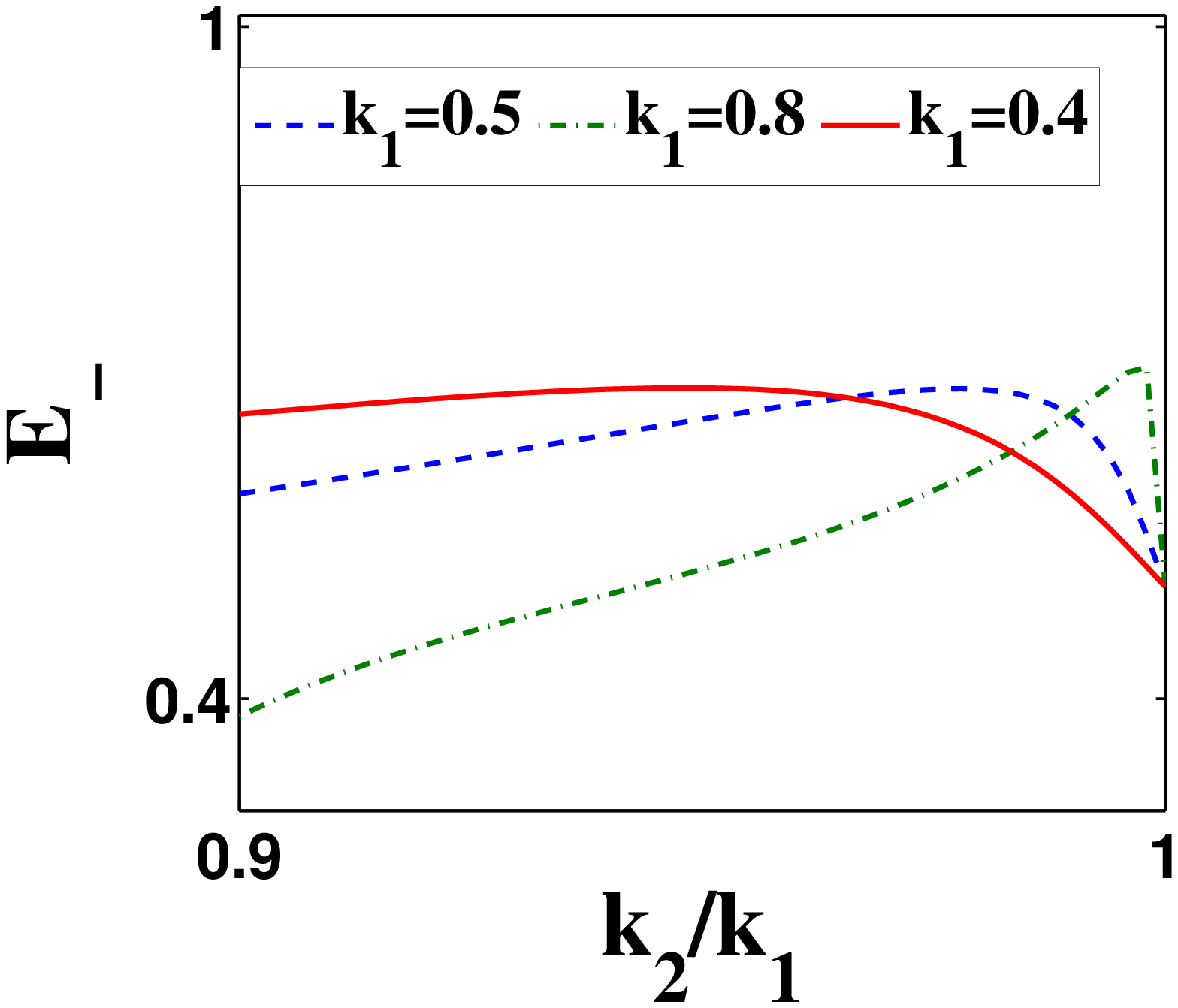}
\vspace{0cm} {\center\footnotesize\hspace{0cm}\textbf{(b)}}\\
\figcaption{(Color Online) Plots of $E_-$ with $k_2/k_1$ varying from 0.9 to 1.0
for two groups of three values of $k_1$. The parameters are chosen as $q=1$,
$\sigma=0.1$, $\Delta=0$, and $x_1/v_1 = x_2/v_2 = 20$. (a) $v_1
=v_2=1$; (b) $v_1=v_2=0.2$. }\label{D}
\end{minipage}
\\[\intextsep]
\\[\intextsep]
\begin{minipage}{\textwidth}
\renewcommand{\captionlabeldelim}{.\,}
\renewcommand{\figurename}{Figs.\,}
\renewcommand{\captionfont}{}
\renewcommand{\captionlabelfont}{}
\vspace{-1cm}\centering
\includegraphics[scale=0.65]{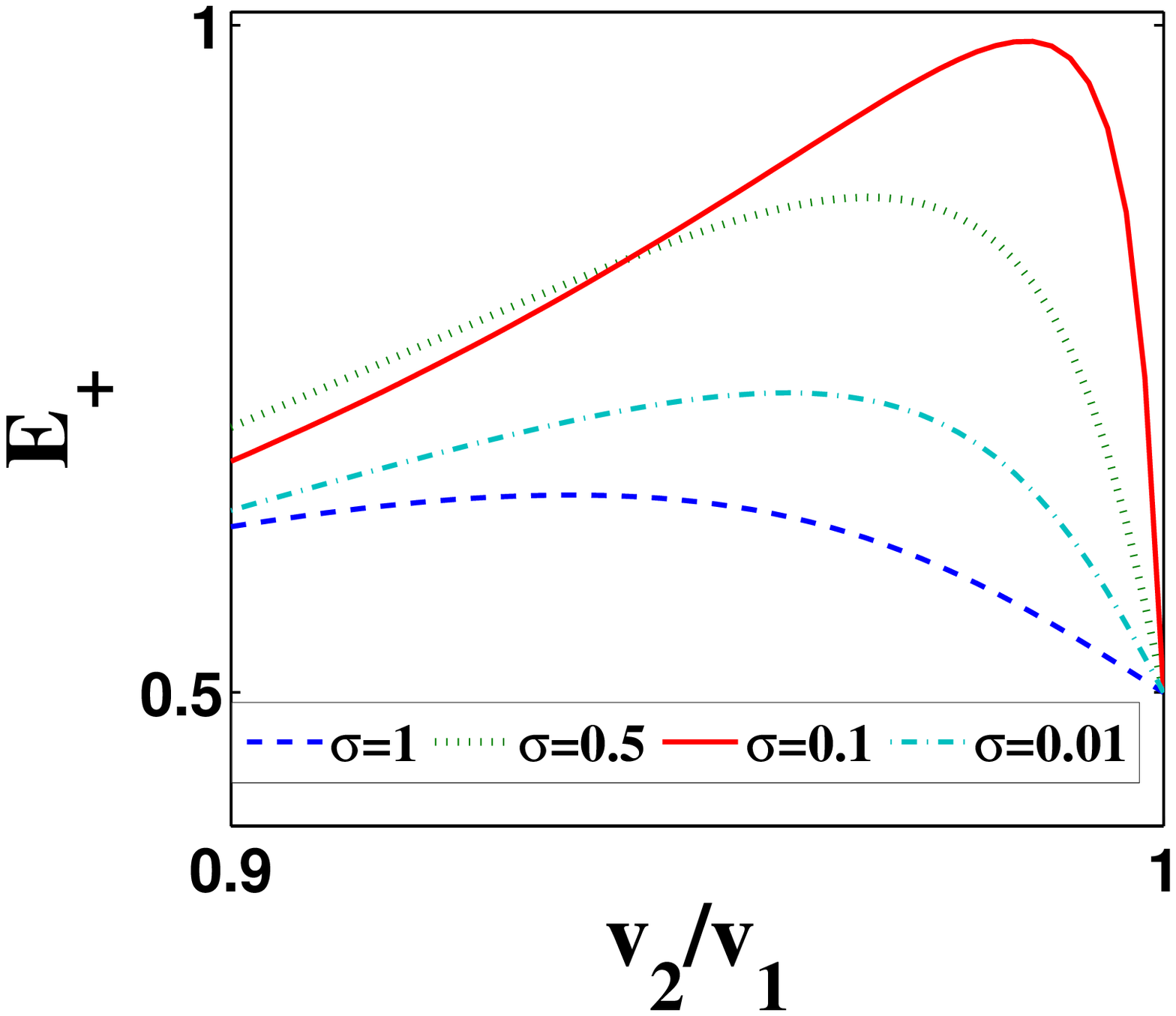}
\vspace{0cm} {\center\footnotesize\hspace{0cm}\textbf{(a)}}\\
\includegraphics[scale=0.65]{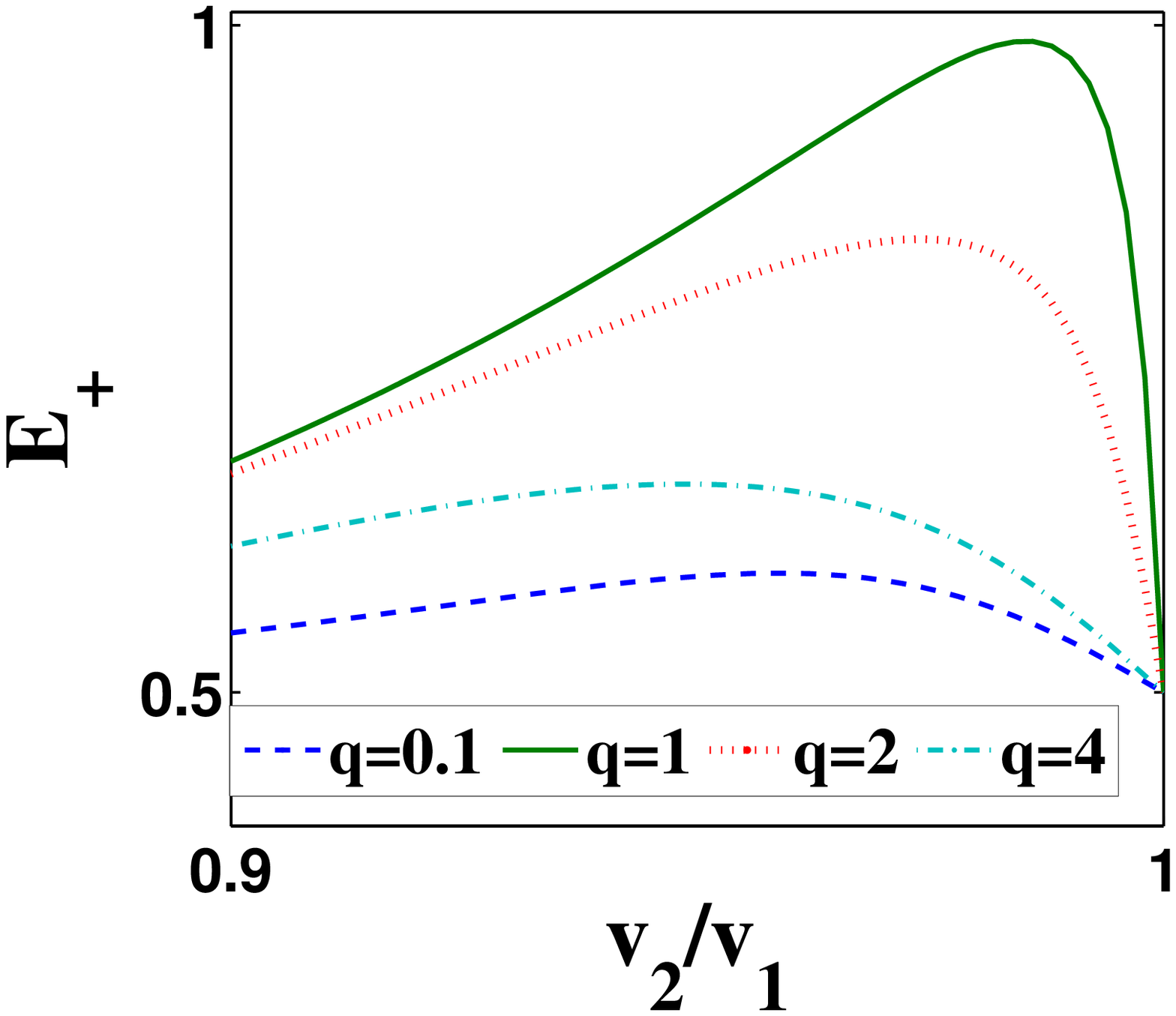}
\vspace{0cm} {\center\footnotesize\hspace{0cm}\textbf{(b)}}\\
\figcaption{(Color Online) (a) Plots of $E_+$ with $v_2/v_1$ varying from 0.9 to 1.0
for four values of $\sigma$ ($q=1$). (b) Plots of $E_+$ with $v_2/v_1$ varying from 0.9 to 1.0
for four values of $q$ ($\sigma=0.1$).  The parameters are chosen as
$k_1=k_2=1$, $v_1=1$, $\Delta=0$, and $x_1/v_1 = x_2/v_2 = 20$. }\label{E}
\end{minipage}
\\[\intextsep]
\\[\intextsep]

\begin{minipage}{\textwidth}
\renewcommand{\captionlabeldelim}{.\,}
\renewcommand{\figurename}{Figs.\,}
\renewcommand{\captionfont}{}
\renewcommand{\captionlabelfont}{}
\vspace{2cm}\centering
\includegraphics[scale=0.65]{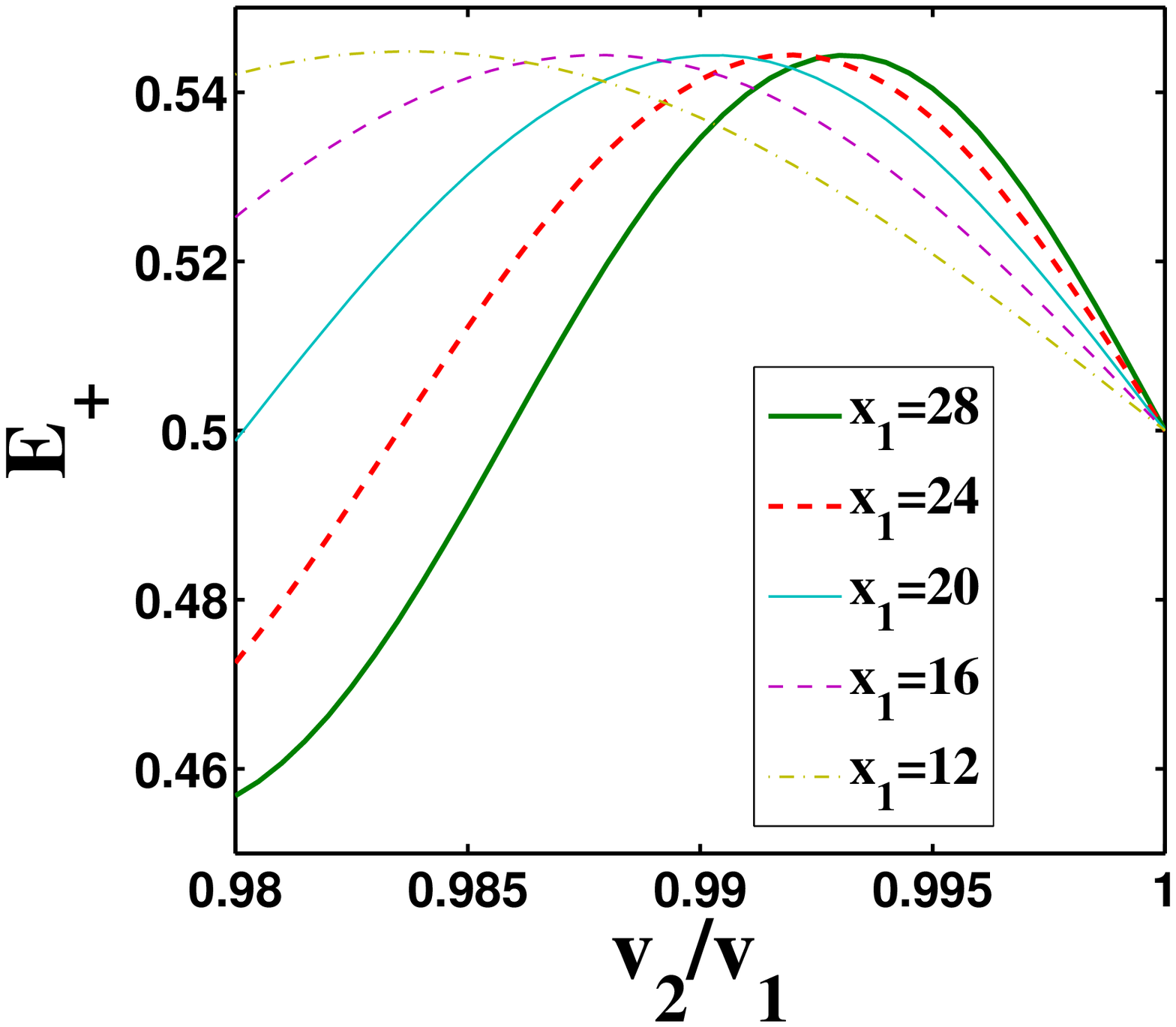}
\vspace{0cm} {\center\footnotesize\hspace{0cm}\textbf{(a)}}\\
\includegraphics[scale=0.65]{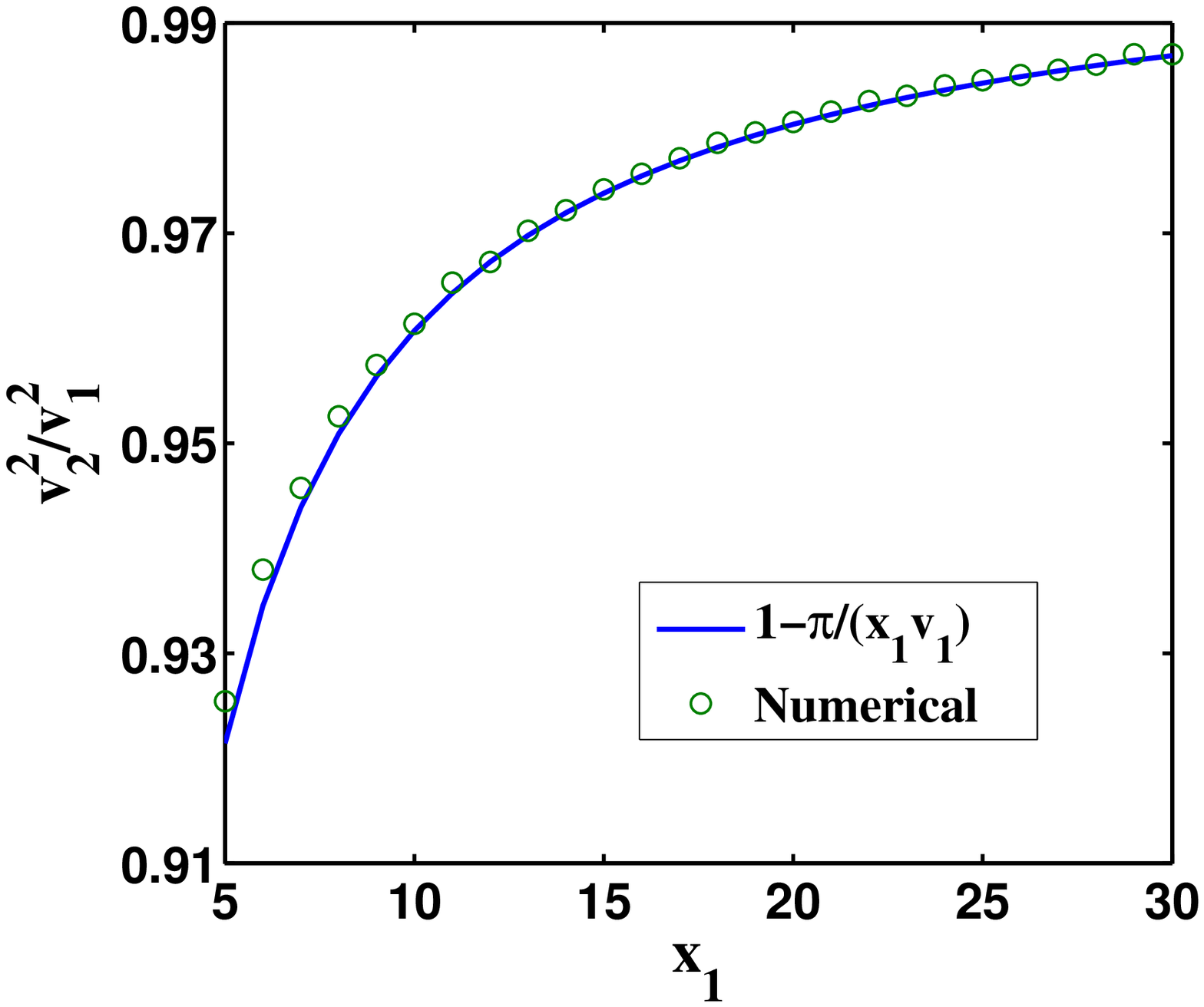}
\vspace{0cm} {\center\footnotesize\hspace{0cm}\textbf{(b)}}\\
\figcaption{(Color Online) (a) Plots of $E_{+}$ as a function of 
$v_2/v_1$ for 5 distinct values of $x_1$. The 
remaining parameters are $q=1$, $\sigma=0.1$, $k_1=k_2=1$, and $v_1=8$.
(b) The point of optimal asymmetry 
quantified by $(v_2/v_1)^2$ is shown as a function
of $x_1$ together with the corresponding theoretical prediction
of our heuristic argument; see the discussion of the last paragraph
of section III. Notice that these panels are
constructed for a large value of speed where, as is illustrated,
this argument is quantitatively valid.}
\label{fignew}
\end{minipage}

\end{document}